\documentclass[
reprint,
superscriptaddress,
amsmath,amssymb,
aps,
prl,
]{revtex4-2}

\usepackage{graphicx}
\usepackage{dcolumn}
\usepackage{bm}
\usepackage{gensymb}
\usepackage{url}
\usepackage{multirow}

\usepackage{xcolor}

\begin{document}

\title{ Orthogonal thermal noise and transmission signals: A new coherent perfect absorption's feature }

\author{Douglas Oña}
\affiliation{%
 Department of Electrical, Electronic and Communications Engineering, Institute of Smart Cities (ISC), Public University of Navarre (UPNA), 31006 Pamplona, Spain
}%

\author{Angel Ortega-Gomez}
\affiliation{%
 Department of Electrical, Electronic and Communications Engineering, Institute of Smart Cities (ISC), Public University of Navarre (UPNA), 31006 Pamplona, Spain
}%

\author{Osmery Hernández}
\affiliation{%
 Department of Electrical, Electronic and Communications Engineering, Institute of Smart Cities (ISC), Public University of Navarre (UPNA), 31006 Pamplona, Spain
}%

\author{Iñigo Liberal}
\thanks{Corresponding author: inigo.liberal@unavarra.es}%
\affiliation{%
 Department of Electrical, Electronic and Communications Engineering, Institute of Smart Cities (ISC), Public University of Navarre (UPNA), 31006 Pamplona, Spain
}%

\begin{abstract}

Coherent perfect absorption (CPA) is an interference process associated with the zeros of the scattering matrix that enables light-with-light interactions in linear systems, of interest for optical computing, data processing and sensing. However, the noise properties of CPA remain relatively unexplored. Here, we demonstrate that CPA thermal noise signals exhibit a unique property: they are orthogonal to the signals transmitted through the network. In turn, such property enables a variety of thermal noise management effects, such as the physical separability of thermal noise and transmitted signals, and ``externally lossless” networks that internally host radiative heat transfer processes. We believe that our results provide a new perspective on the many CPA technologies currently under development. 

\end{abstract}

\maketitle

\section{Introduction}

Coherent perfect absorption (CPA) is an interference process where the proper combination of several input signals cancels out all output signals, leading to complete absorption \cite{chong2010coherent,zhang2012controlling,baranov2017coherent}. Mathematically, CPA stems from the analytical properties of the scattering matrix $\mathbf{S}$, whereby adding loss to a system moves the zeros of $\mathbf{S}$ down onto the real axis. CPA can also be interpreted as the time-reversed process of lasing as threshold \cite{chong2010coherent}, where the signals are outgoing rather than ingoing, and where the poles of $\mathbf{S}$ are moved upwards onto the real axis by the addition of gain \cite{Ge2010steady}. 

CPA has been experimentally demonstrated in multiple configurations including planar slabs \cite{Wan2011time}, metasurfaces \cite{zhu2016coherent}, graphene layers \cite{rao2014coherent}, integrated photonics \cite{espinosa2018coherent,Bruck2013plasmonic,Zanotto2017coherent,Hernandez2022quantum}, surface plasmon polariton gratings \cite{jung2015} and epsilon-near-zero (ENZ) media \cite{luo2018coherent}. CPA in the single photon regime has been observed with plasmonic metasurfaces \cite{Roger2015coherent,Vetlugin2019coherent,vetlugin2021coherent,Hernandez2022generalized}. Recently, CPA has been demonstrated in the context of disordered media in analogy with random lasers \cite{pichler2019random}, coupled microresonators supporting CPA exceptional points \cite{Wang2021coherent}, chaotic CPA modes in large resonators supporting thousands of resonances \cite{jiang2023coherent}, and optical cavities with massively degenerated CPA modes \cite{Slobodkin2022}. CPA have also facilitated the control waves within complex environments with the use of reconfigurable metasurfaces \cite{sol2022meta,erb2023control,cuesta2022coherent}. Beyond optical systems, CPA has been performed with matter waves in Bose-Einstein condensates \cite{Mullers2018coherent} and sound waves in acoustic setups \cite{meng2017acoustic}.

Because CPA critically depends on the coexistence of multiple input signals, it effectively enables light-with-light interactions in a linear system \cite{zhang2012controlling}. Thus, it finds natural applications in interferometry \cite{Wan2011time}, all-optical data processing \cite{Papaioannou2016all}, sensing, as well as strengthening and dynamically controlling absorption-based  processes such as photocurrent generation \cite{liew2016coherent} and photoluminiscence \cite{pirruccio2016coherent}.

Here, we theoretically demonstrate that CPA modes enable orthogonal channels for thermal noise and transmission signals. We find that because of the algebraic properties of CPA modes, the heat radiated by an optical network and the signals transmitted through it occupy orthogonal vector spaces. The same property allows for the internal exchange of heat withing an optical network, while simultaneously confining all thermal signals withing the network and remaining transparent to all their input optical modes. We believe these results bring a fresh perspective to the physics of CPA, and it could be applied to the many CPA technologies that are currently under investigation \cite{chong2010coherent,zhang2012controlling,baranov2017coherent,Ge2010steady,Wan2011time,zhu2016coherent,rao2014coherent,
espinosa2018coherent,Bruck2013plasmonic,Zanotto2017coherent,Hernandez2022quantum, 
jung2015,luo2018coherent,Roger2015coherent,Vetlugin2019coherent,vetlugin2021coherent,Hernandez2022generalized,pichler2019random,Wang2021coherent,jiang2023coherent,Slobodkin2022,
sol2022meta,erb2023control,cuesta2022coherent,Mullers2018coherent,meng2017acoustic,Papaioannou2016all,liew2016coherent,pirruccio2016coherent}.
It might also open new opportunities in radiative heat and energy management \cite{Fan2022photonics,baranov2019nanophotonic,Joulain2005surface}, as well as in the design of nanophotonic thermal engines \cite{Gelbwaser2021near,Tsurimaki2022moving,Park2022,Wan2011time,zhu2016coherent}.

\begin{figure}[!t]
  \centering
    \includegraphics[width=0.75\columnwidth]{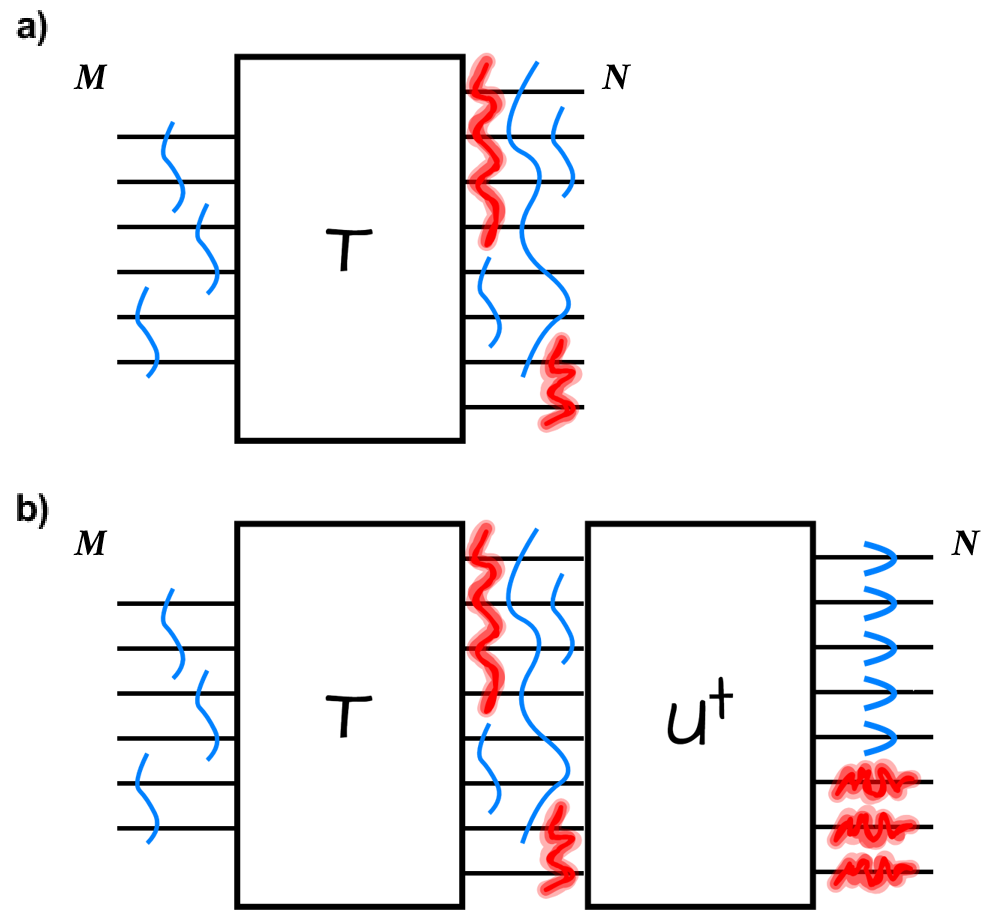}
  \caption{a) Schematic of a network of $M$ left ports and $N$ right ports. b) Physical separation of CPA thermal noise and transmitted signals by means of a unitary network.}
  \label{fig:1}
\end{figure}

To guide our thoughts, we consider an optical network with $M$ left and $N$ right ports, with $N>M$ (see Fig. \ref{fig:1}(a)). We assume that all ports are matched, so that there are no back reflections. The vectors of input and output signals, $\mathbf{a}$ and $\mathbf{b}$ respectively, are related throught a $\mathbf{S}\in\mathbb{C}^{\left(M+N\right)\times\left(M+N\right)}$ scattering matrix $\mathbf{b}=\mathbf{S}\mathbf{a}$. We label the $M+N$ ports such that the first $M$ ports correspond to the ports on the left of the network, while the subsequent $N$ ports are the ports on the right (see Supplementary Material). Then, if all input ports are matched and the network is reciprocal, the scattering matrix can be written as follows
\begin{equation}
\mathbf{S}=\left[\begin{array}{cc}
\mathbf{0}_{M\times M} & \mathbf{T}^{T}\\
\mathbf{T} & \mathbf{0}_{N\times N}
\end{array}\right]
\label{eq:S_general}
\end{equation}

\noindent where $\mathbf{T}\in\mathbb{C}^{N\times M}$ is the transmission matrix charaterizing the left-to-right transmission of signals through the network. For any linear network, $\mathbf{T}$ admits a singular value decomposition (SVD) \citep{Miller2012all,Hernandez2022generalized} $\mathbf{T}=\mathbf{U}\mathbf{D}\mathbf{V}^{\dagger}$, where $\mathbf{U}\in\mathbb{C}^{N\times N}$ is a unitary matrix $\mathbf{U}\mathbf{U}^{\dagger}=\mathbf{I}_{N}$ providing a basis for the signals outgoing the $N$ ports on the right, and $\mathbf{V}\in\mathbb{C}^{M\times M}$ is a unitary matrix $\mathbf{V}\mathbf{V}^{\dagger}=\mathbf{I}_{M}$ providing a basis for the signals incoming through the $M$ left ports. $\mathbf{D}\in\mathbb{R}^{+N\times M}$ is a matrix of singular values with the following structure:
\begin{equation}
\mathbf{D}=\left[\begin{array}{c}
\mathbf{D}_{T}\\
\cdots\\
\mathbf{0}_{N-M\times M}
\end{array}\right]
\end{equation}

\noindent with $\mathbf{D}_{T}=\mathrm{diag}\left\{ d_{1},\ldots,d_{M}\right\} $ and $d_{n}\in\mathbb{R}^{+}\,\,\,\forall n$ are the singular values. It is also interesting to structure $\mathbf{U}$ as follows
\begin{equation}
\mathbf{U}=\left[\begin{array}{ccc}
\mathbf{U}_{T} & \vdots & \mathbf{U}_{0}\end{array}\right]
\end{equation}

\noindent highlighting that while the columns of $\mathbf{U}$ span the complete vectorial space in the right output ports, with dimension $N$, the transmission signals that can be excited from the left are restricted to the spaced spanned by the columns of $\mathbf{U}_{T}\in\mathbb{C}^{N\times M}$ with dimension $M$. The space spanned by the columns of $\mathbf{U}_{0}\in\mathbb{C}^{N\times N-M}$ with dimension $N-M$ are inaccessible to the signals. Inspecting the SVD of $\mathbf{S}$ (see Supplementary Material) shows that these inaccessible $N-M$ modes correspond to CPA modes when the system is excited from the right. Therefore, the inaccessibility of these modes can be understood as a consequence from the fact that reciprocity forbids the excitation of CPA modes. In addition, if any of the remaining singular values equals zero, $d_{n}=0$, we would have additional CPA modes, also inaccessible to the transmitted signals. In general, any reflectionless network with an asymmetric number of ports contains at least $N-M$ channels that are inaccessible to the transmitted signals, and the signal processing tasks performed by the network.

At the same time, any lossy network at temperature $T$ emits thermal radiation, which can be characterized by the noise correlation matrix \cite{haus2001optimum,zhu2013temporal,miller2017universal}
\begin{equation}
\left\langle \mathbf{n}\mathbf{n}^{\dagger}\right\rangle =\left(\mathbf{I}-\mathbf{S}\mathbf{S}^{\dagger}\right)N_{T}
\end{equation}

\noindent where $N_{T}=\frac{1}{2\pi}\,\hslash\omega\,\left(e^{\frac{\hslash\omega}{k_{B}T}}-1\right)^{-1}$ is the blackbody energy spectrum. After a number of algebraic manipulations (see Supplementary Material), we find that the correlation matrix admits the following eigendecomposition
\begin{equation}
\left\langle \mathbf{n}\mathbf{n}^{\dagger}\right\rangle 
=\mathbf{P}\,\mathbf{D}_{N}\,\mathbf{P}^{\dagger}\,\,N_{T}
\end{equation}

\noindent where $\mathbf{P}\mathbf{P}^{\dagger}=\mathbf{I}_{N+M}$ is a unitary matrix, constructed from the $\mathbf{V}$ and $\mathbf{U}$ matrices as follows
\begin{equation}
\mathbf{P}=\left[\begin{array}{cc}
\mathbf{V}^{*} & \mathbf{0}_{M\times N}\\
\mathbf{0}_{N\times M} & \mathbf{U}
\end{array}\right]
\end{equation}

\noindent and
\begin{equation}
\mathbf{D}_{N}=\mathrm{diag}\{1-d_{1}^{2},\ldots,1-d_{M}^{2},1-d_{1}^{2},\ldots,1-d_{M}^{2},\mathbf{1}_{1\times N-M}\}
\label{eq:D_N}
\end{equation}

The first $2M$ diagonal entries in (\ref{eq:D_N}) show the existence of thermal signals for each $n^{th}$ channel for which $d_{n}<\text{1}$. For these channels, the transmitted power is reduced by a factor $d_{n}^{2}$, while a $\left(1-d_{n}^{2}\right)N_{T}$ noise power is added into the channel. Thus, thermal noise and transmitted signals overlap in the same channel when $d_{n}<\text{1}$, leading to known reduction of the signal-to-noise ratio (SNR) by a lossy device.
On the other hand, the last $N-M$ diagonal entries in (\ref{eq:D_N}) equal one, meaning that the $N-M$ CPA modes inaccessible to the transmitted signals also act as perfect blackbody thermal emitters. However, the additional noise introduced by CPA thermal signals occupies an orthogonal space to the space of transmitted signals. For a system in which $d_{n}=\text{1}\,\,\,\forall n$, all the thermal emission from the device and the transmitted signals would occupy orthogonal spaces.

\textit{\textbf{-Physical separability:}} A direct consequence from the fact that thermal CPA and transmitted signals occupy orthogonal channels is that they can be physically separated. Again, for a system with only either transparent or CPA channels, i.e., $d_{n}=\text{1}\,\,\,\forall n$, all thermal noise and transmitted signals occupy orthogonal channels. Thus, if we change the basis of the thermal noise signals to $\mathbf{n}_{p}=\mathbf{P}^{\dagger}\mathbf{n}$, the correlation matrix is diagonalized as $\left\langle \mathbf{n}_{p}\mathbf{n}_{p}^{\dagger}\right\rangle =\mathrm{diag}\{\mathbf{0}_{1\times 2M},\mathbf{1}_{1\times N-M}\}\,N_{T}$. Therefore, using a detector that selectively separates such vectorial spaces, it would be possible to simultaneously transmit information signals and radiative heat without a reduction of the SNR.

In addition, since $d_{n}=\text{1}\,\,\,\forall n$ and $N>M$, all noise signals exit through the right. In this case, the physical separation can be implemented by adding a network implementing the unitary transformation $\mathbf{U}^{\dagger}$ (see Fig. \ref{fig:1}(b)). Such transformation changes the transfer matrix to $\mathbf{T}_{aux}=\mathbf{I}_N\mathbf{D}\mathbf{V}^{\dagger}$, where the basis for the output signals is given by the identity matrix, i.e., each output mode of the SVD decomposition is assigned to a physical port. We note that there are known algorithms for the design of optical networks implementing any arbitrary unitary transformation \cite{miller2013self,pai2019matrix}.

\begin{figure}[h]
  \centering
    \includegraphics[width=\columnwidth]{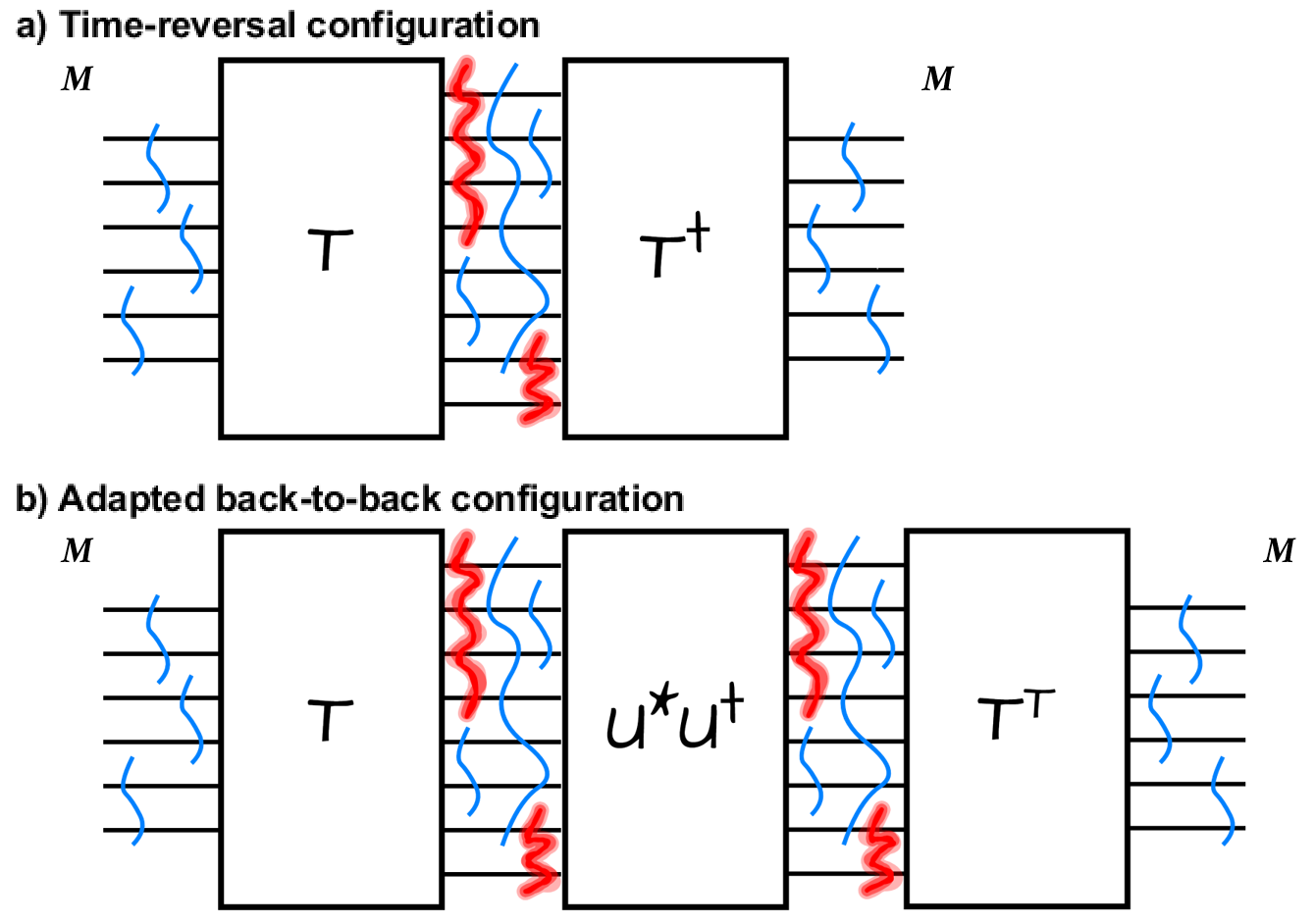}
  \caption{a) Time-reversal and b) adaptated back-to-back configurations}
  \label{fig:2}
\end{figure}

\textit{\textbf{-Time-reversal processing:}} The algebraic properties of CPA thermal noise signals also produce nontrivial heat transfer phenomena within time-reversal processing \cite{tanter2000time}. To this end, let us assume that the aforementioned network is connected with a network characterized by transmission matrix $\mathbf{T}^{\dagger}$  (see Fig.\,\ref{fig:2}(a)), which effectively implements a time-reversal operator \cite{tanter2000time}. If the transmission matrix is unitary, $\mathbf{T}^{\dagger}\mathbf{T}=\mathbf{I}_M$, the network reproduces the input signals. However, one cannot generally expect $\mathbf{T}^{\dagger}$ to be unitary for lossy networks, which generally fail to produce time-reversal operators. Here, it is important to remark that any reciprocal network exhibits time-reversal symmetry, and the input signals can be recovered by running the time-reversal of the output signals. What we state here is that lossy networks cannot reproduce a time-reversal operator, where the transmission through the network is compensated by a linear time-reversal operator.

In general, for the time-reversal configuration (TR) depicted in Fig.\,\ref{fig:2}(a), the resulting network is an $M\times M$ network, characterized by scattering matrix $\mathbf{S}_{TR}=\left[\begin{array}{ccccc}
\mathbf{0}_{M\times M} & \mathbf{T}_{TR}^{T} & ; & \mathbf{T}_{TR} & \mathbf{0}_{M\times M}\end{array}\right]$. In this configuration, the transfer matrix is given by $\mathbf{T}_{TR}=\mathbf{T}^{\dagger}\mathbf{T}=\mathbf{V}\mathbf{D}_{T}^{2}\mathbf{V}^{\dagger}$, which is now a positive semidefinite Hermitian matrix. Only if $d_n=1\,\,\,\forall n$ we have that $\mathbf{D}_{T}^{2}=\mathbf{I}_M$ and the network recovers time-reversal processing. Interesting, we note that even if $d_n=1\,\,\,\forall n$ the network is still lossy, since it has $N-M$ CPA channels, but it nevertheless performs time-reversal processing. 

Again, the thermal noise signals exiting the network are characterized by the noise correlation matrix, which is given by (see Supplementary Material)
\begin{equation}
\left\langle \mathbf{n}\mathbf{n}^{\dagger}\right\rangle =
\mathbf{P}_{TR}\,\mathbf{D}_{TR}\,\mathbf{P}_{TR}^{\dagger}\,\,N_{T}
\label{eq:N_TR}
\end{equation}

\noindent with
\begin{equation}
\mathbf{P}_{TR}=\left[\begin{array}{cc}
\mathbf{V}^{*} & \mathbf{0}_{M\times M}\\
\mathbf{0}_{M\times M} & \mathbf{V}
\end{array}\right]
\label{eq:P_TR}
\end{equation}

\noindent and
\begin{equation}
\mathbf{D}_{TR}=\mathrm{diag}\{1-d_{1}^{4},\ldots,1-d_{M}^{4},1-d_{1}^{4},\ldots,1-d_{M}^{4}\}
\label{eq:D_NR}
\end{equation}

It is clear from (\ref{eq:D_NR}) that the thermal signals of the $N-M$ CPA channels do not exit the network and are perfectly contained within the device. In fact, if the $M$ modes are fully transparent, i.e., $d_{n}=1\,\,\forall n$, no thermal signal would be observed outside the network. At the same time, there are CPA thermal signals flowing in the network, which lead to observable phenomena. If the two sub-networks are at temperature $T_1$ and $T_2$, respectively, the net flux of radiative thermal between the two sub-networks would be  $P_{\mathrm{th}}=\left(N-M\right)(N_{T_1}-N_{T_2})$. Therefore, the properties of CPA modes enable the thermal radiative transmission of energy within a network, while it remains noiseless and transparent from the outside.

\textit{\textbf{-Back-to-back configuration:}} It must be noted that time-reversal is not a simple back-to-back configuration, since the latter is characterized by a transmission matrix $\mathbf{T}^{T}$ instead of $\mathbf{T}^{\dagger}$. Therefore, unless all elements of $\mathbf{T}$ are real, the back-to-back configuration presents a different response from the time-reversal configuration.  For example, a back-to-back configuration does not trap all CPA thermal noise signals within the network.  

However, the trapping of CPA thermal noise signals can be recovered by introducing a matching network that performs the unitary transformation $\mathbf{U}_c=\mathbf{U}^*\mathbf{U}^\dagger$ (see Fig\,\ref{fig:2}(b)). Such network changes the basis of the output and input channels of $\mathbf{T}$ and $\mathbf{T}^{T}$, respectively, so that they match, in what can be called an \textbf{adapted back-to-back (aBB)} configuration. Hence, the property of containing CPA thermal noise signals is not exclusive to time-reversal processing networks. 

For such aBB configuration, the resulting network is an $M\times M$ network characterized by the scattering matrix $\mathbf{S}_{aBB}=\left[\begin{array}{ccccc}
\mathbf{0}_{M\times M} & \mathbf{T}_{aBB}^{T} & ; & \mathbf{T}_{aBB} & \mathbf{0}_{M\times M}\end{array}\right]$, where the transfer matrix is $\mathbf{T}_{aBB}=\mathbf{T}^{T}\mathbf{U}_{c}\mathbf{T}=\mathbf{V}^{*}\mathbf{D}_{T}^{2}\mathbf{V}^{\dagger}$ (see Supplementary Material). In this case, if $d_n=1\,\,\,\forall n$, we have that the transmission matrix reduces to $\mathbf{T}_{aBB}=\mathbf{V}^{*}\mathbf{V}^{\dagger}$, which is not necessarily the identity matrix. Thus, we find that the property of heat exchange mediated by CPA thermal channels, is not restricted to time-reversal processing and it is compatible with more arbitrary unitary transformations. 

\begin{figure*}[!t]
  \centering
  \includegraphics[width=\textwidth]{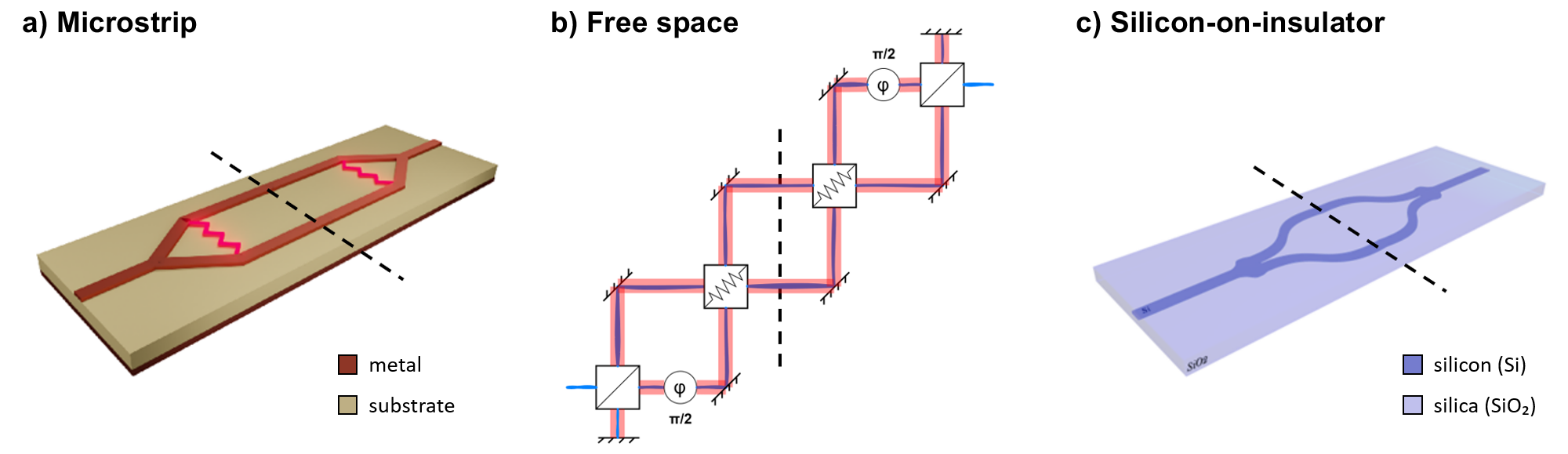}
  \caption{\textbf{Examples of time-reversal configurations} based on two Wilkinson power dividers (WPDs) implemented in a) microstrip, b) free space and, c) silicon-on-insulator (SOI) technologies.}
  \label{fig:3}
\end{figure*}

For the aBB configuration, the correlation matrix is given by (see Supplementary Material)
\begin{equation}
\left\langle \mathbf{n}\mathbf{n}^{\dagger}\right\rangle =
\mathbf{P}_{aBB}\,\mathbf{D}_{aBB}\,\mathbf{P}_{aBB}^{\dagger}\,\,N_{T}
\label{eq:N_aBB}
\end{equation}

\noindent with
\begin{equation}
\mathbf{P}_{aBB}=\left[\begin{array}{cc}
\mathbf{V}^{*} & \mathbf{0}_{M\times M}\\
\mathbf{0}_{M\times M} & \mathbf{V}^{*}
\end{array}\right]
\label{eq:P_aBB}
\end{equation}

\noindent and
\begin{equation}
\mathbf{D}_{aBB}=\mathbf{D}_{TR}
\label{eq:D_NB}
\end{equation}

Equations (\ref{eq:N_aBB})-(\ref{eq:D_NB}) confirm that this configuration also ensures the full containment of CPA thermal noise signals within the network.

\textit{\textbf{-Examples of applicability:}} The theoretical effects described above apply to general CPA networks, and it is expected that they can be observed in many of the CPA technologies currently under investigation \cite{chong2010coherent,zhang2012controlling,baranov2017coherent,Ge2010steady,Wan2011time,zhu2016coherent,rao2014coherent,
espinosa2018coherent,Bruck2013plasmonic,Zanotto2017coherent,Hernandez2022quantum, 
jung2015,luo2018coherent,Roger2015coherent,Vetlugin2019coherent,vetlugin2021coherent,Hernandez2022generalized,pichler2019random,Wang2021coherent,jiang2023coherent,Slobodkin2022,
sol2022meta,erb2023control,cuesta2022coherent,Mullers2018coherent,meng2017acoustic,Papaioannou2016all,liew2016coherent,pirruccio2016coherent}. Next, we present a few examples on how these effects emerge on some popular CPA devices. First, we consider a Wilkinson power divider (WPD) \cite{pozar2011microwave}, i.e., a $1\times 2$ device featuring a single CPA channel that has been studied within the context of quantum state transformations \cite{Hernandez2022quantum}. The scattering matrix of a WPD is given by:

\begin{equation}
\mathbf{S}_{\rm WPD}=\frac{1}{\sqrt{2}}\begin{bmatrix}
0 & 1 & 1\\ 
1 & 0 & 0\\ 
1 & 0 & 0
\end{bmatrix}
\label{eq:S_WPD}
\end{equation}

By comparing Eqs.\,(\ref{eq:S_general}) and (\ref{eq:S_WPD}) we find that a WPD is an example of a reflectionless network with an asymmetric number of ports, whose transfer matrix is $\mathbf{T}_{WPD}=1/\sqrt{2}\,\,\,\mathbf{1}_{2\times 1}$. In addition, all elements of the transmission matrix are all real, so that $\mathbf{T}^{T}=\mathbf{T}^{\dagger}=1$, and the time-reversal (TR) and back-to-back (BB) configurations are identical. Therefore, in both TR and BB configurations the total network matrix ($\mathbf{S}_{T}$) can be expressed as
\begin{equation}
\mathbf{S}_{T}=\begin{bmatrix}
0 & 1\\ 
1 & 0
\end{bmatrix}
\label{eq:S_total}
\end{equation}

It is clear from Eq.\,(\ref{eq:S_total}) that the combined system is perfectly transparent and lossless, as it could be expected from a time-reversal processing configuration. Therefore, WPDs provide an example of lossy CPA networks that remain perfectly transparent to electromagnetic waves via time-reversal processing, while allowing for the simultaneous transfer of heat between two sub-networks. 

This effect could be observed in a variety of technological platforms as illustrated in Fig.\,\ref{fig:3}. First, WPDs have been traditionally implemented at microwave frequencies by using microstrip lines and resistors (see Fig.\,\ref{fig:3}(a)). They are routinely used in microwave networks, being a common component of beamforming networks \cite{ruchenkov2018implementation,albuquerque2015implementation,9769044} and amplification stages \cite{1353514,214619}. In addition, generalizations to WPDs with arbitrary number of output ports and splitting ratios have been demonstrated \cite{5546961}. Interestingly, the same behavior could be reproduced in other technologies. For example, Fig.\,\ref{fig:3}(b) depicts an free-space optical setup where the WPDs are implemented with a combination of beamsplitters, lossy beamsplitters and mirrors (See supplementary material for a theoretical description of the network). On the other hand, Fig.\,\ref{fig:3}(c) shows a system implemented in integrated optics through silicon-on-insulator (SOI) technology. In this case, modified Y-branches can be used as WPDs, where the loss can be introduced either by plasmonic materials or radiative losses \cite{Hernandez2022quantum}.

\begin{figure*}[!t]
  \centering
  \includegraphics[width=\textwidth]{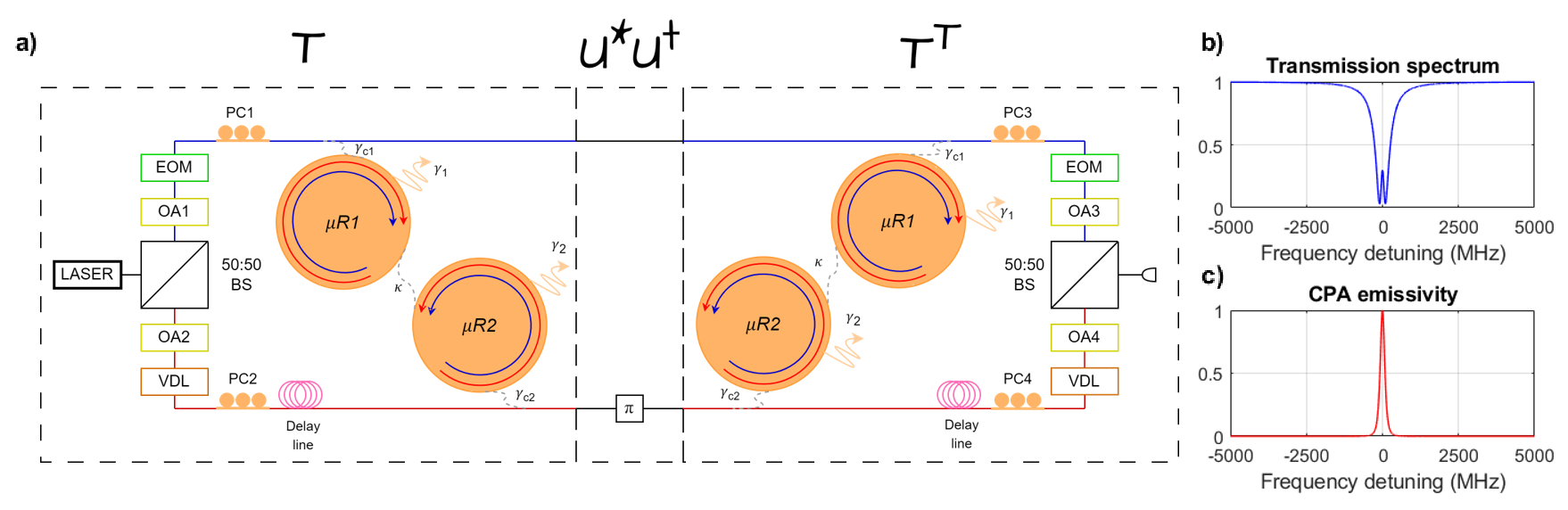}
  \caption{\textbf{Example of an adapted back-to-back (aBB) configuration} based on the system presented in \cite{Wang2021coherent}. b) Transmitted signal and c) internal heat transfer spectra. These results were obtained for the following parameters: $\gamma_{1}=64.782\,MHz$, $\gamma_{2}=242.93\,MHz$, $\gamma_{c1}=184.63\,MHz$ and $\gamma_{c2}=123.09\,MHz$, enforcing a generic CPA EP.}
  \label{fig:4}
\end{figure*}

Finally, we present a example that does not operate in a TR configuration and presents filtering effects with a nontrivial dispersion profile, based on the recently studied CPA at an exceptional point (EP) \cite{Wang2021coherent}. The system (see Fig.\,\ref{fig:4}(a)), consists of two silica microtoroidal resonators, coupled to two single-mode fiber conical waveguides. It is a very general platform that leads to different singular responses as a function of the system parameters. We focus on the ``Generic CPA EP", for which two zeros converge on the real axis, forming an EP with CPA properties. That is to say, two eigenvalues of the system are simulateneously zero and two eigenvectors coalesce to $1/\sqrt{2}\,[1\, , \, i]$. 

As detailed in the Supplementary Material, an adapted back-to-back configuration can be constructed by using a unitary matrix $\mathbf{U}_{c}=\mathbf{U}_{G}^{*}\mathbf{U}_{G}^{\dagger}=\mathrm{diag}\{1,-1\}$, which can be easily implemented with a delay line (shown as a $\pi$ block in Fig.\,\ref{fig:4}(a)). Fig.\,\ref{fig:4}(b) shows the transmission through the system when internal phase shifters are tuned to excite $1/\sqrt{2}\,[1\, , \, -i]$ signals orthogonal to the EP eigenvalue. The transmission is characterized by an absorbing doublet. At the same time, the internal emissivity associated with the EP eigenvector $1/\sqrt{2}\,[1\, , \, i]$ presents a single peak at the EP frequency (see Fig.\,\ref{fig:4}(c)). At this point, CPA signals are transferred between both resonator networks, with unit efficiency, without disturbing the transmitted signals. If the resonator loss were dissipative, this mechanism would allow for heat transfer between both resonator systems. If the losses were predominantly radiative, such channel would allow the coupling of external signals through the resonators. In general, this example shows that it is possible to implement filtering with a nontrivial dispersion profile, while simultaneously enabling heat transfer between two subnetworks that does not increase the externally observable noise. 

Our results demonstrate that CPA thermal noise signals have a singular property: they are orthogonal to the signals transmitted through the network. This property can be understood as a form of spatial coherence, which enables the physical separability of CPA thermal noise signals via networks implementing unitary transformations. In addition, such property can be harnessed in time-reversal and adaptative back-to-back configurations, enabling heat transfer channels that remain confined within the network, and thus do not increase the externally observable noise. In general, we believe that our results highlight the nontrivial thermal noise properties of CPA, which remain relatively unexplored. CPA wave phenomena are being currently investigated in a large number of technological platforms \cite{chong2010coherent,zhang2012controlling,baranov2017coherent,Ge2010steady,Wan2011time,zhu2016coherent,rao2014coherent,
espinosa2018coherent,Bruck2013plasmonic,Zanotto2017coherent,Hernandez2022quantum, 
jung2015,luo2018coherent,Roger2015coherent,Vetlugin2019coherent,vetlugin2021coherent,Hernandez2022generalized,pichler2019random,Wang2021coherent,jiang2023coherent,Slobodkin2022,
sol2022meta,erb2023control,cuesta2022coherent,Mullers2018coherent,meng2017acoustic,Papaioannou2016all,liew2016coherent,pirruccio2016coherent}, and we believe that our results present a new perspective in which to look and reexamine CPA systems.

\bibliography{library}

\end{document}